**OPEN FORUM**

# Artificial intelligence in cyber physical systems

Petar Radanliev[1] · David De Roure[1] · Max Van Kleek[2] · Omar Santos[3] · Uchenna Ani[4]

**Abstract**
This article conducts a literature review of current and future challenges in the use of artificial intelligence (AI) in cyber physical systems. The literature review is focused on identifying a conceptual framework for increasing resilience with AI through automation supporting both, a technical and human level. The methodology applied resembled a literature review and taxonomic analysis of complex internet of things (IoT) interconnected and coupled cyber physical systems. There is an increased attention on propositions on models, infrastructures and frameworks of IoT in both academic and technical papers. These reports and publications frequently represent a juxtaposition of other related systems and technologies (e.g. Industrial Internet of Things, Cyber Physical Systems, Industry 4.0 etc.). We review academic and industry papers published between 2010 and 2020. The results determine a new hierarchical cascading conceptual framework for analysing the evolution of AI decision-making in cyber physical systems. We argue that such evolution is *inevitable and autonomous* because of the increased integration of connected devices (IoT) in cyber physical systems. To support this argument, taxonomic methodology is adapted and applied for transparency and justifications of concepts selection decisions through building summary maps that are applied for designing the hierarchical cascading conceptual framework.

## 1 Introduction

Artificial intelligence (AI) is already changing our economy and society, and the increased AI decision making has triggered debated on the potential harms and the need to make AI decision making more transparent (de Fine Licht et al. 2020, forthcoming). Even with our current technological progress, self-building technologies are possible (Kammerer 2020, forthcoming). Cognitive architectures representing truly intelligent human-like performance, that includes '*motivation, emotion, personality, and other relevant aspects,*' are also possible (Sun 2020). Such findings trigger concerns on the creation of collective '*Borg–eye and the We–I*' subjects, by merging the desires of many subjects, e.g. though wearable connected devices, into a collective (Liberati 2020).

This articulates research questions on how the increased AI decision making is changing our economy and society, and how we can make AI decision making more transparent. These questions contribute to furthering the discussion of this paper, especially in view of *intrusive self-building technologies* that represent truly intelligent

✉ Petar Radanliev
  petar.radanliev@oerc.ox.ac.uk

  David De Roure
  david.deroure@oerc.ox.ac.uk

  Max Van Kleek
  max.van.kleek@cs.ox.ac.uk

  Omar Santos
  osantos@cisco.com

  Uchenna Ani
  u.ani@ucl.ac.uk

1   Engineering Science Department, Oxford e-Research Centre, University of Oxford, 7 Keble Road, Oxford OX1 3QG, England, UK
2   Department of Computer Science, University of Oxford, Oxford, England, UK
3   Cisco Research Centre, Research Triangle Park, NC, USA
4   Faculty of Engineering Science, STEaPP, University College London, London, England, UK

human-like performance, triggering the creation of collective from desires of many subjects.

One such intrusive technology is the Industrial Internet of Things (IIoT). Internet of Things (IoT) technology has become of considerable academic, government, and industry interest in recent years. The IIoT can be explained as the use of internet of things technologies to improve manufacturing and industrial processes. The IIoT term is closely related to the term Industry 4.0 (I4.0), which represents at the same time: a paradigm shift in industrial production, a generic designation for sets of strategic initiatives to boost national industries, a technical term to relate to new emerging business assets, processes and services, and a brand to mark a very particular historical and social period.

Through reviewing a considerable academic, government and industry literature, specific research questions emerge from the research gaps that the review has identified. There is a significant gap in current research on how the integration of complex and interconnected internet of things (IoT), coupled in cyber physical systems (CPS), triggers *inevitable and autonomous* evolution of artificial cognition. The literature review and taxonomic analysis consider the significance of these research gaps in the discussion on how technological advancement results with the *inevitable and autonomous* evolution of artificial cognition in complex, coupled and interconnected socio-technical systems.

One example for Artificial Intelligence (AI) working in combination with internet of things (IoT) devices is the Tesla car. The car uses Artificial Intelligence (AI) to determine road conditions, optimal speed, weather, and to predict pedestrians' and cars' movement. Another example, in the context of Covid-19, is the use of smart buildings. While the internet of things (IoT) can be used as sensors for switching on lights and opening doors, in combination with Artificial Intelligence (AI), it could also be used for predicting optimal time for heating or cooling the building. In the future, Artificial Intelligence (AI) in cyber physical systems (CPS) would include health and biomedical monitoring, robotics systems, intelligent edge devices, among many other functions, and be used to correct natural disasters, human errors, or malicious actions, etc.

Hence, this is exercise is important, because with the increased number of internet of things (IoT) connected devices, the role of cyber physical systems (CPS) has changed and evolved. With the added element of Industrial Internet of Things (IIoT) increasing productivity, efficiency and economic benefits, and the changing role of Artificial Intelligence (AI) used for the creation of this new economic benefits, the current five levels of cyber physical system architecture seems obsolete. With considerations of these new technologies, we focus on determining a new CPS architecture.

In this article, we refer to Artificial Intelligence (AI) not only as a technology for reasoning, planning, learning, and processing, but we also refer to the ability to move and manipulate objects. This relates or research on Artificial Intelligence (AI) with Cyber Physical Systems (CPS). By Cyber Physical Systems (CPS), we refer to computer–human networks, controlling physical processes, where physical processes affect computations and vice versa. One modern version of Cyber Physical Systems is the Internet of Things (IoT). The Internet of Things (IoT) is one step forward in the advancement of AI in machines and represents a system of interrelated computing devices, capable of operating without human-to-human or human-to-computer interaction. The Industrial Internet of Things (IIoT) in this study refers to sensors and other devices networked with industrial applications, enabling data collection, exchange, and analysis, with the objective for increase in productivity, efficiency and economic benefits.

The review of such systems in this paper includes the advancements in Cyber Physical Systems (CPS), the Internet of Things (IoT) in relation to Artificial Intelligence (AI) autonomous evolution in Industry 4.0 (I4.0). In this context, we propose the term CPS-IoT to refer to the integration of cyber physical attributes into Industrial Internet of Things (IIoT) systems. This integration includes advances in real-time processing, sensing, and actuation between IIoT systems and physical domains and provides capabilities for system analysis of the cyber and physical structures involved. We, therefore, focus here on artificial cognition, defined as the artificial intelligence in networked connection of people, processes, data, and things. Therefore, artificial intelligence in this article represents a more inclusive and encompassing concept that consolidates the cyber physical attributes of IIoT with the social aspects of the environment in which this technology is deployed and reflects the future cognitive makeup of IIoT/I4.0. The term artificial cognition in the context of this article is used to discuss effect from the evolving IoT services and social networks of I4.0.

This article is structured as follows: Our methodology is described in Sect. 2. In Sect. 3, we discuss the findings drawn from the literature review including contributions and gaps that form artificial cognition in CPS. Section 4 produces a taxonomy for management techniques and their significance to the discussion on artificial cognition in I4.0. A Discussion section and a Conclusion section synthesise our findings and ends the article.

## 2 Methods

The methods applied in this study consist of systematic literature review, taxonomies derived and follows existing research studies on this topic that apply literature review with taxonomy (Milano et al. 2020), in pursuit of narratives (O'Hara 2020). Academic literature and practical studies are consulted intensively to discuss the IoT technologies and their relation to the I4.0. While the mainstream academic literature offers limited insights regarding existing and emerging cognitive developments, we use summary maps to showcase recent developments in this field.

Our rationale is that—as the landscape of artificial cognition develops and changes very quickly—merely relying on journal publications provides too narrow a view of the present situation. We used the analytical target cascading, combined with the grounded theory approach (Glaser and Strauss 1967), to construct a conceptual cascading model for the future integration of cognition in the I4.0. These models then inform a qualitative empirical study for the new cognitive feedback mechanism approach. The chosen method for conducting systematic literature review represented the following: (1) searching established journal databases and updating the findings with cross checking with google scholar search engine; (2) creating a table of search terms and article inclusion criteria such as relevance, peer review, data of publication (less than 10 years), and design of studies. (3) we also considered ethical issues in relations to how data was obtained, reported, and protected. For example, we did not include any non-peer-reviewed studies that were critical of different nations or organisations. We also did not include any literature where data sources were not included. For example, studies that claim individual company and/or nation CPS or IoT performance was better (e.g. Huawei vs Ericson vs Nokia) were not included if the data were not included in the study, or if we were unable to verify the results.

## 3 Literature review on cyber risk analytics and artificial intelligence

The literature review is focused on identifying the most prominent concepts present in current models, infrastructures and frameworks, from over 90 academic, government and industry papers, reports, and technical notes, published predominately between 2010 and 2020. In our search for data records, we used predominately Google Scholar and the Web of Science Core Collection. For selecting the academic literature, we found Google Scholar more flexible when adding more search terms. For example, when adding multiple terms in the Web of Science Core Collection, with the Boolean: AND, the search results are limited. We searched for TOPIC: (artificial intelligence) AND TOPIC: (industrial internet of things) AND TOPIC: (internet of things) AND TOPIC: (cyber physical systems) AND TOPIC: (industry 4.0). This search on the Web of Science Core Collection produced only 25 data records. If only one of the Booleans: AND is changed to OR, then the data records change to hundreds of thousands, but its relevance to the correlated topics diminishes, and focus is placed on the one topic searched with the Boolean: OR. We repeated the same search with Google Scholar, with all topics TOPIC: (artificial intelligence) AND TOPIC: (industrial internet of things) AND TOPIC: (internet of things) AND TOPIC: (cyber physical systems) AND TOPIC: (industry 4.0). The same search on Google Scholar produced 20,700 data records. Hence, to ensure the relevance to all of the topics we investigated, of our selected data records, we used both the Web of Science Core Collection and Google Scholar, but since the number of articles was much greater on Google Scholar, we used predominately the Google Scholar search engine for analysing the greater volume of data records. Since both databases contain articles from the same journals, and Google Scholar search engine is more effective in search queries on many topics, using Booleans, we considered this as valid argument for selecting the most relevant data records.

Since the existing CPS architecture that we reviewed and tried to update was published in 2015, we tried to include literature predominately from the time period between 2015 and 2020. However, some of the most important literature from 2010 to 2020 is also included, and for inclusiveness, a very few articles from before 2010 are included in the review. Considering the purpose of this review was to update our understanding of CPS architecture, we did not conduct a historic analysis of all relevant literature. Instead, we considered that the CPS architecture from 2015 included knowledge from historic literature, and our aim was to update that knowledge with the most recent findings on CPS architecture.

Concepts that are recognised as most prominent are categorised following the grounded theory approach for categorising emerging concepts (Glaser and Strauss 1967). This process is detailed in the 'Methods' chapter. As a result of following the (Glaser and Strauss 1967) research approach arguing that *'all you see is data'*, the categorising of most prominent concepts identified from over 90 different sources, the emerging categories of concepts are diverse in research nature. Throughout the paper, the reader meets terms related to: (1) economic potential; (2) cognitive design; (3) risk engineering; (4) correlation effect; (5) cognitive feedback; (6) 'unrecognised and outdated data. These six terms are just examples of the plethora of different terms and concepts that emerge from our literature review on the topic of cyber physical system architecture. We categorised these terms and concepts, and redesigned the exiting five levels of cyber physical system architecture—or 5C (Fig. 1).

The grounded theory method is applied to categorise these diverse terms and concepts to the existing architecture that comprises five levels of cyber physical systems or 5C (Fig. 1). The grounded theory approach is used to categorise these new terms and concepts, emerging form the literature review, and organised into cascading hierarchies of actions (in Table 1), presented as summary maps. The importance of these diverse concepts and the relationship between seemingly unrelated concepts, is what coheres to the design of the proposed hierarchical cascading approach (in Fig. 2).

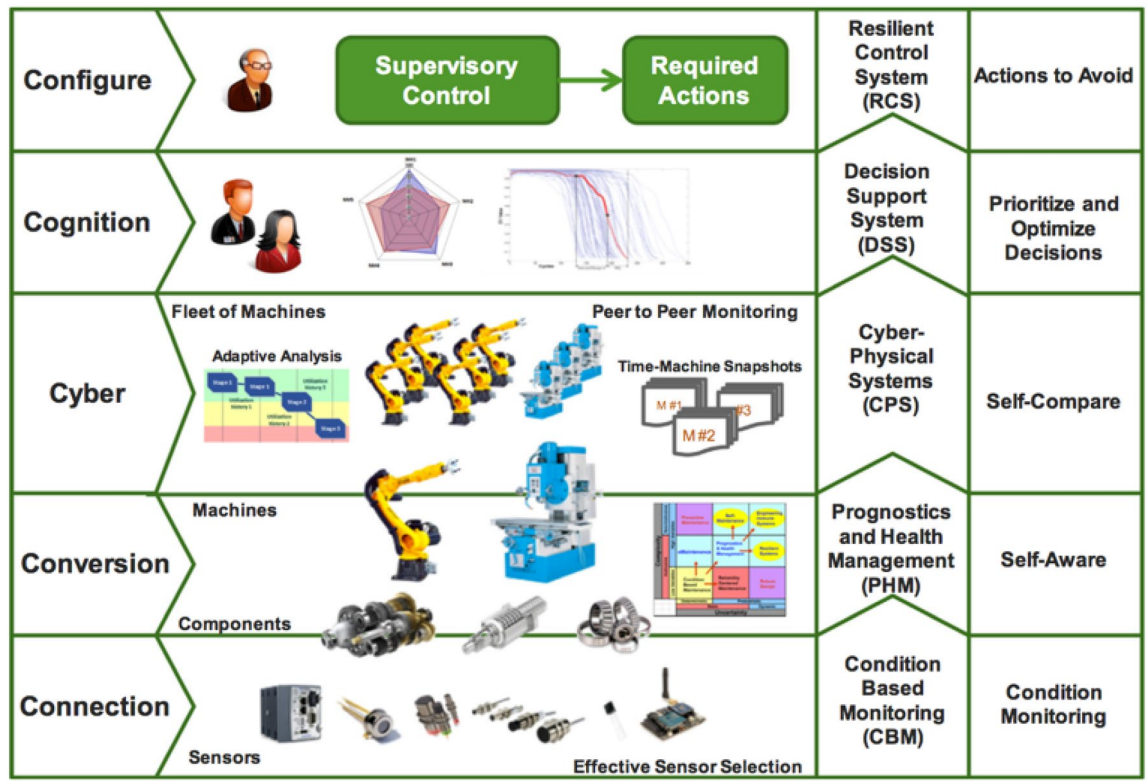

**Fig. 1** The 5 levels cyber physical system architecture—commonly referred to as 5C architecture

The complexity of the literature coherent design becomes more explicit with examples that are presented throughout the paper. The examples place the paper within the experiential and cultural practice of engineering. Here we present one explicit example of how the research questions that are drawn from the literature review are then included to drive new finding and contributions on the identified gaps in existing literature. The first example is used to drive conceptual and theoretical underpinnings of the research gap. This example from literature derives findings that the exact economic impact of cognitive CPS infrastructure remains to be determined (Leitão et al. 2016) although cognitive CPS systems will represent a large percentage of future ICT application in industry (Marwedel and Engel 2016). This situation presented in this example requires a new approach for integrating the physical and cyber subsystems of cognitive CPS. The new approach needs to provide an overall understanding of the design, development, and evolution of cognition in CPS, and needs to integrate theories of artificial intelligence, control of physical systems, as well as their interaction with humans.

Such approach is especially needed for not only developing nations that lack an I4.0 strategies, but also for more developed countries—such as the UK and USA. The UK has been ranked as the overall global cyber superpower followed by the US (Allen and Hamilton 2014). It is also reported that the UK and US are strongly protected to withstand digital infrastructure cyber-attacks, which is crucial in developing a resilient digital economy. However, in the index quantifying industrial applications in digital infrastructure key sectors, the UK drops down to the 5th place and the US to the 3rd place. This seems to be partly due to the UK and US lagging behind other countries in terms of harnessing economic value from the I4.0 (Allen and Hamilton 2014). This could be caused by the lack of cognitive abilities in the Internet of Things (IoT) deployment (Radanliev et al. 2020a).

The literature review continues with identifying, categorising and relating emerging concepts to the conceptual and theoretical underpinnings of the arguments that cohere to the conceptual framework design.

Table 1 Summary map—table of technologies that drive artificial cognition in CPS

| Taxonomy of key elements that drive AI | |
|---|---|
| CPS—cognitive communities | |
| Cyber physical systems | CPS |
| Internet of everything | IoE |
| 5 level CPS architecture | 5C |
| Agent-oriented architecture | AoA |
| Object-oriented architecture | OoA |
| Cloud optimised virtual object architecture | VOA |
| Virtual engineering objects | VEO |
| Virtual engineering processes | VEP |
| Model-driven manufacturing systems | MDMS |
| Service oriented architecture | SoA |
| Dynamic intelligent swamps | DIS |
| CPS—cognitive processes | |
| Connected devices and networks | CDN |
| Compiling for advanced analytics | CfAA |
| Business processes and services | BPS |
| Cloud distributed process planning | DPP |
| Physical and human networks | PHN |
| CPS—cognitive societies | |
| Internet of things | IoT |
| Web of things | WoT |
| Social manufacturing | SM |
| Internet of people | IoP |
| Internet of services | IoS |
| Systems of systems | SoS |
| CPS—cognitive platforms | |
| Internet protocol version 6 | IPv6 |
| Internet-based system and service platforms | ISP |
| Model-based development platforms | MBDP |
| Knowledge development and applications | KDoA |
| Real-time distribution | RtD |

## 3.1 Values and risks from intrusive autonomous self-building connected technologies (IoT, edge computing) in cyber physical systems

One of the main drives for artificial intelligence in cyber physical systems is value creation. Our society is driven by social-economic values. Organisational goals are always based on some form of values. For example, governmental and non-governmental sectors are driven by the development of societal values. Private organisations are often driven by economic values. One of the main drives for value creation is the emerging new data streams that enable understanding of new events in real-time, and predicting future events. This new and emerging data come at volumes that only AI can process with low-latency. Since this value emerges from cyber physical systems, it becomes inevitable that autonomous AI will evolve in economic and societal decision making.

This process is already in motion, triggered by the enormous economic potential for hyper-connected economy. Literature recognises that important future business opportunities lay in the networking potential of digital economy (Nicolescu et al. 2018). The infrastructure for smart manufacturing technology could create large cost savings for manufacturers (Anderson 2016) and enable faster development of economies of scale (Brettel et al. 2016). Industrial Internet, or 'Industry 4.0,' supports a finer granularity and control to meet individual customer requirements, creates value opportunities (Hermann et al. 2016; Shafiq et al. 2015; Stock and Seliger 2016; Wang et al. 2016), increases resource productivity, and provides flexibility in business processes (Hussain 2017). The integration of cognitive cyber-physical capabilities into IIoT arguably requires a new process for integrating physical and cyber subsystems—including an overall understanding of the cognitive design, development, and evolution of CPS and IIoT. Gaining such understanding may require consolidation of IIoT theories for control of physical systems and the interaction between humans and CPS (Marwedel and Engel 2016; Roure et al. 2019; Banks 2019).

On the other hand, the US National Institute of Standards and Technology (US NIST) deliberately stays away from formalising any process model in this space (Barrett et al. 2017; NIST 2018). Instead, their recent Framework for Cyber Physical Systems proposes sets of artefacts and activities that could be considered by organisations in the deployment of CPS. These proposals are the result of formal ontologies of digital artefacts and their interactions with the exterior world. The US NIST identifies three main views on CPS that encompass identified responsibilities in the systems engineering process: conceptualisation, realisation, and assurance. Each of these three views corresponds to fundamental processes in the life of cognitive CPS, respectively: (1) Models of CPS (design), (2) the CPS itself (implementation), and (3) CPS Assurance (validation). The trade-offs between different instantiations of these processes as well as between critical aspects such as Security, Safety, Business, and Privacy need to be understood. In this context, Risk Engineering is proposed as an activity embedded in the design, development, and lifecycle of the future CPS and IoT systems (Radanliev et al. 2020b). This assumes that cyber risk is just one instantiation of risk for an organisation or product and, therefore, should be subject to the higher processes of compliance and regulation in each domain. Building on this understanding of risk, a cognitive feedback approach is needed for formalising compositional ways to reason about cyber risks in an I4.0 context. For example, what we could do to understand and measure the systemic IIoT risk is to create a requirement for automatic sharing of cyber-attacks data records between IIoT supply chains. If IoT connected devices are reporting the standalone risks of a sole company, this would enable supply chain participants to understand and differentiate between stand alone and systemic cyber risk. However, when IoT connected devices start reporting on standalone risks of a sole company, this could create duplicate data records, collection of irrelevant data records, and many other complications. Hence, the cyber-attack reporting needs to include an element of cognition, possibly in the fog computing layer, because it would be challenging to implement cognition in the edge computing systems.

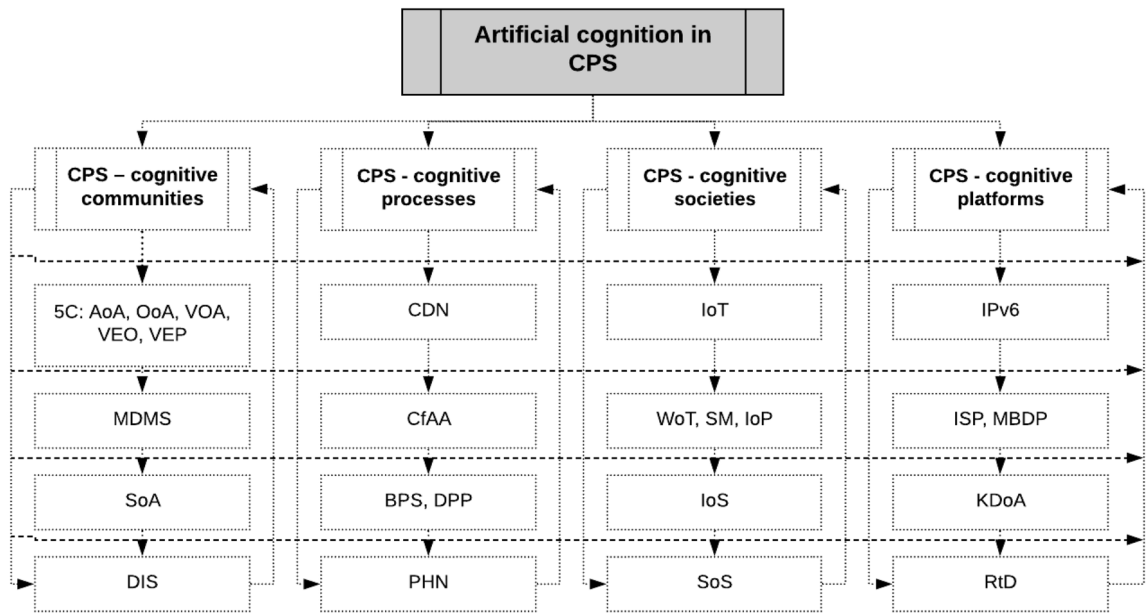

**Fig. 2** Hierarchical cascading framework design, describing how artificial intelligence is evolving in CPS

### 3.2 Argument for cognitive analytics

The arguments for cognitive feedback approach emerge from the inherent risk in integrating the physical with the cyber world, where cyber risk environment is constantly changing (Radanliev et al. 2018), and estimated loss of cybercrime varies greatly (Biener et al. 2014; DiMase et al. 2015). The real impact of cyber risk remains unknown (Shackelford 2016), mainly due to lack of suitable probabilistic data and lack of a universal, standardised impact assessment framework (Radanliev et al. 2020b; Koch and Rodosek 2016). To develop such a framework, accumulated risk needs to be quantified in real-time and shared across technology platforms (Ruan 2017). This requires a dynamic understanding of the network risk. In addition, new risk elements that require cognitive analytics also need to be quantified, such as intellectual property of digital information (Anthonysamy et al. 2017) and the impact of media coverage (Tanczer et al. 2018).

### 3.3 Review on existing cyber risk analytics

The Cyber Value at Risk (CvaR) model (World Economic Forum 2015), represents an attempt to understand the economic impact of cyber risk for individual organisations. CVaR provides cyber risk measurement units, value analysis methods related to the cost of different cyber-attacks type (Roumani et al. 2016), and proof of concept methods that are based on data assumptions. Given the lack of data needed to validate the CvaR model, these studies calculate the economic impact based on organisations' 'stand-alone' cyber risk and, therefore, ignore the correlation effect of sharing infrastructure and information and the probability of cascading impacts, which represents a crucial element of I4.0. These limitations of the CvaR model are of great concern, e.g. in sharing cyber risk in critical infrastructure (Zhu et al. 2011). Critical infrastructures are vital for strong digital economies, but issues of synchrony, components failures, and increasing complexity demand development and elaboration of new rigorous CPS methods (Rajkumar et al. 2010). In the absence of a common reference point of cyber risks, existing cyber risk assessment methodologies have led to inconsistencies in measuring risk (Agyepong et al. 2019), which negatively affects the adaptation of I4.0. Assessment of IIoT cyber risk in I4.0 should be based on a system that enables cognitive assessment of the cyber network risk, not only the stand-alone cyber risks (Craggs and Rashid 2017) of a sole company (Radanliev 2014).

## 3.4 Review of financial assessment of cyber risk from CPS

In early literature, existing financial models have been proposed to assess information security investment (Anderson and Moore 2006; Gordon and Loeb 2002; Rodewald and Gus 2005). However, cyber risk covers more elements than information security financial cost, such as brand reputation (Lee et al. 2019a) or intellectual property (Lee et al. 2019b). In terms of modelled economic and financial impact of massive cyber-attacks, additional questions emerge in relation to the impact on public sector, rethinking of business processes, growth in liability risk, and mitigation options (Ruffle et al. 2014). Such economic evaluations trigger a debate between limited economic lifespans of digital assets and value in inheriting 'out of date' data (Tan et al. 2008). In an I4.0 context, cyber risks are not only simply associated with machines and products that store their knowledge and create a virtual living representation in the network (Drath and Horch 2014) but also to the global flows and markets they are part of.

## 4 Taxonomy of management technologies and methodologies on AI-enabled methods

This section redefines the Fig. 1—5C architecture (5 levels of CPS architecture) and creates a taxonomy from the chapter 3—literature review. The taxonomy represents a list of focal points, listed in a summary table (Table 1), for visualising and focusing the direction for a new CPS architecture. To define the contribution from this study, before we present the new cognitive feedback mechanism, we first explain the existing 5C architecture in Fig. 1 as described in (Lee et al. 2015). The purpose of including Fig. 1 was to discuss the weaknesses of the current understanding of CPS architecture.

From Fig. 1, we can see that the current five levels cyber physical system architecture (5C) includes one level for cyber elements. With the rise of connected devices—IoT and IIoT, and AI in human–computer interactions, the cyber level is obsolete, because each level contains various cyber elements. In this study, we seek for improved understanding of CPS architecture and we seek that though a taxonomy of recent literature.

The new cognitive feedback mechanism builds upon the existing recommendations that CPS needs to adapt quickly (Niggemann et al. 2015), to create multi-vendor and modular production systems (Weyer et al. 2015). Requiring understanding of multi-discipline testing (Balaji et al. 2015), system sociology (Dombrowski and Wagner 2014), and social networks (Wan et al. 2015; Roure et al. 2015).

## 4.1 Key technologies for self-adapting system

Before conducting the taxonomic categorisations in the summary tables (see chapter 5) in this final section we compress the rationale for the categorisations (see Table 1). This section also details the four categories, which is one less category than the five levels of CPS presented in Fig. 1. Our CPS architecture does not include the 'cyber' level, which was considered as a separate level in the previous architecture. We argue that cyber is far more than an individual level: we argue that cyber is part of all levels of the CPS architecture.

The academic literature we analysed outlines the evolution of CPS into the more inclusive and encompassing system that brings together people, process, data, and things—making networked connections and transactions more valuable to individuals, organisations, and things. Hence, by applying grounded theory for categorising the literature analysed, the following key feedback management technologies predominated: (a) integration of physical flows, information flows, and financial flows; (b) innovative approaches to managing operational processes; (c) exploiting the industrial digitisation to gain competitiveness; (d) and utilisation of Big Data to improve the efficiency of production and services. From the extensive literature reviewed on this topic, the requirements for cognitive feedback are categorised in Table 1 as: follows domain communities, processes, societies, and platforms. These domains represent how the changing roles of innovation, production, logistics, and the service processes require CPS advancements in the following: (a) domain communities; (b) internet-based system and service platforms; (c) business processes and services, and (d) dynamic real-time data from physical and human networks (perceived as data from intelligent swamps). This introduces the approach used for the taxonomic categorisations and the summary tables in chapter 5.

## 5 Summary of the taxonomic analysis: building summary maps

Although we described the process in the previous section, we wanted to explain further that this section—chapter 5—is summarising the findings from the literature review in chapter 4 and categorises the emerging terms and concepts into actions and activities, presented as hierarchical cascades of activities in a summary map (Table 1). The summary map in Table 1 is the first step in building a new theory and improving the current five levels of CPS architecture with a new and more up-to-date architecture. Before presenting the summary map, we briefly discuss the categorisations as described in the previous section (domain communities, processes, societies, and platforms.) and refer to their origin—with references from the literature review.

### 5.1 Taxonomic categorisations for advancing the existing 5C architecture

The domain communities, processes, societies, and platforms, are expanded into (1) domain communities; (2) internet-based system and service platforms; (3) business processes and services, and (4) dynamic real-time data from physical and human networks.

*Domain communities* include the following: Agent-oriented Architecture (Ribeiro et al. 2010), Object-oriented Architecture (Thramboulidis 2015), Cloud optimised Virtual Object Architecture (Giordano et al. 2016), supported with Virtual Engineering Objects and Virtual Engineering Processes with Internet Protocol version 6 (IPv6) connected devices and networks (Wahlster et al. 2013).

*Internet-based system and service platforms* (La and Kim 2010) are used to model CPS through the Web of Things (Dillon et al. 2011), with compiling of data, processes, devices, and systems for cognitive analytics and connection to cognitive model-driven (robot-in-the-loop) manufacturing systems (Jensen et al. 2011; Shi et al. 2011; Wang et al. 2014). Internet-based system and service platforms can be used to promote model-based development platforms, such as behaviour modelling of robotic systems, e.g. Automata (Ringert et al. 2015). Internet-based systems and service platforms can enable the development of social manufacturing and interconnect with the Internet of People to create CPS collaborative communities (Lee et al. 2014).

*Business processes and services* need to be interconnected into industrial value chains to integrate machine information into decision making and be connected to the Internet of Services for service oriented CPS architecture (Wang et al. 2015) and Cloud distributed planning manufacturing. Business processes and services in CPS can also promote knowledge development of business areas and applications.

*Dynamic real-time data from physical and human networks (perceived as dynamic intelligent swamps)* of modules connected to physical and human networks, can operate as systems of systems, and can act as mechanisms for real-time distribution (Kang et al. 2012) and feedback directly from users and markets.

### 5.2 Summary map of emerging terms and concepts—presented as actions and activities

The categories of key elements for artificial cognition in CPS are presented in Table 1. The relationships of these elements to CPS is grouped with the grounded theory into the following categories: CPS—cognitive communities, CPS—cognitive processes, CPS—cognitive societies and CPS—cognitive platforms. These categories and the synergies between the elements lead to artificial cognition in CPS for self-aware process are categorised in Table 1. The taxonomic analysis of the literature reviewed is applied to structure closely related concepts higher and looser relationships lower within each category in the Table 1 summary maps. These communities, processes, societies, and platforms emerged from categorising the literature review findings. The taxonomic interpretation of the relationships between these concepts is built upon the literature findings and represent the backbone of theoretical development and its understanding of interconnected concepts in this paper. We created the taxonomic categorisations in Table 1 to seek improvement and update of the existing CPS architecture (see Fig. 1). In the taxonomy, we relate the merging concepts to the original concept in Fig. 1, but we do not include the 'cyber' layer. We considered cyber to be an integral part of all layers in CPS architecture. Hence, the taxonomy in Table 1 contains four levels of CPS architecture.

In brief, the summary table (Table 1) can also be seen as multiple cascading hierarchies of actions—found in literature as terms and concepts. The six terms mentioned in the introduction of chapter 3—literature review, are dissected in greater detail, with more specific focus on presenting actions and activities, not desired objectives. For example, from the six terms, we used the first term '(1) economic potential;' and in the literature review, we investigated for actions and activities that are related to this term. In the summary table (Table 1), we can see new terms and concepts, e.g. Business processes and services; Model-driven manufacturing systems; etc. The wording in these terms and concepts is structured in a more actionable form. For example, the term (1) 'economic potential' does not provide any guidance on how this economic potential can be achieved. We just discovered that 'economic potential' was strongly present in literature on cyber physical system architecture. So we used this term as one of the six guidance terms in the introduction of chapter 3. But in the summary table (Table 1), we can see these terms as actionable concepts, e.g. 'Model-driven manufacturing systems' that explain what needs to be done to reach the 'economic potential'. The summary table (Table 1) presents multiple cascading hierarchies of actions that are used in the design of the cascading hierarchy framework in (in Fig. 2). Instead of presenting these actions and activities as categories related to the six terms, we used the recommendations from the literature where we found the related actions and activities. We wanted to determine if these six terms are true representation of all the terms and concepts in literature, or was a different structure more relevant.

The summary maps in Table 1 confirm that a notion of artificial cognition in CPS goes beyond machine to machine (M2M) (Wan et al. 2013; Stojmenovic 2014), and beyond the proposed 3 level CPS, which are (1) services, (2) cloud, and (3) physical object layers. Artificial cognition in CPS also goes beyond the existing knowledge of 5C architecture (as seen in Fig. 1). When artificial cognition in CPS is combined with intelligent manufacturing equipment (Posada et al. 2015), then a new set of communities, processes, societies, and platforms (categorised in Table 1) emerge. When combined, these new machineries represent intrusive self-building technologies, triggering an inevitable and autonomous evolution of artificial cognition in CPS.

This evolution goes beyond the existing description of 5C architecture (in Fig. 1). The new description of artificial cognition in CPS (as seen in Table 1) is based on the integration of artificial intelligence (AI), machine learning, the cloud, and IoT, creating systems of machines capable of interacting with humans (Carruthers 2014). For example, the application of behaviour economics into CPS already enables market speculation on human behaviour (Rutter 2015), and even neuromarketing (Lewis and Brigder 2004), to determine consumer purchasing behaviour. We can expect to see autonomous CPS adopting the use of these methods to predetermine human behaviour.

Technologies described in Table 1 that would enable artificial cognition in CPS include software defined networks (Kirkpatrick 2013) and software-defined storage (Ouyang et al. 2014), built upon the following: protocols and enterprise grade cloud hosting; AI, machine learning, and data analytics (Kambatla et al. 2014; Pan et al. 2015); and mesh networks and peer-to-peer connectivity (Wark et al. 2007). Without cognitive risk analytics, the embedded control of CPS is creating security and risk management vulnerabilities from integrating less secured systems, triggering questions regarding risk management and liability for breaches and damages (Boyes et al. 2018). Without cognitive risk analytics, many other technical challenges can be foreseen in the CPS vital domains—especially in the design, construction, and verification of CPS (Anthi et al. 2019).

# 6 From summary maps to conceptual framework

In this section, we use the hierarchical cascading method, with categorical coding (Radanliev 2014), to build a conceptual framework based on the findings in the summary map Table 1. Our aim was not to confirm that the embodiment of AI in the IIoT is leading to a transformation in AI; we consider that as a given—postulate from the beginning of this study. Our aim in this section was to present advancement to the current 5 levels CPS architecture (5C) as presented in Fig. 1 and to integrate the plethora of emerging concepts from our literature review, which are not included in the current 5C architecture—(in Fig. 1).

The summary maps in Table 1 should be seen from a conceptual standpoint, and not from engineering perspective on the definition of terms. If seen on a standalone bases, the summary maps in Table 1 could be seen as concepts that represent a diverse set of different terms. From reading the summary maps categorisations in Table 1, the Internet Protocol v6 is categorised as a platform, while from an engineering perspective IPv6 is a networking protocol. There are multiple categorisations of this type. To reduce the categories and themes in our pursuit of deeper understanding of these categories, the grounded theory approach used the Pugh-controlled convergence and, in the process, themes are associated with the 'best fit' categories. The rationale for this categorisation is as follows: Protocol (e.g. the Internet Protocol v6) is the official procedure or system of rules governing the communication or activities of programs and/or industries. Platform on the other hand refers to the technologies that are used as a base upon which other applications, processes or technologies are developed. A CPS in the context of this categorisation is a platform, while the languages it uses to communicate (e.g. IPv6) with software are the protocol.

Further clarification as why such categorisations have been made by applying the Pugh-controlled convergence to reduce the number of categories is that we can consider a platform as a software, while protocol is more like a theory, or theoretical model which a platform can be based on. In the interest of keeping the cascading hierarchy design to a level that can easily be understood, the presented categorisations have been associated in abbreviated form in Table 2.

Table 2 Emerging 4 levels CPS architecture

| Artificial cognition in CPS | | | |
| --- | --- | --- | --- |
| CPS—cognitive communities | CPS—cognitive processes | CPS—cognitive societies | CPS—cognitive platforms |
| 5C: AoA, OoA, VOA, VEO, VEP | CDN | IoT | IPv6 |
| MDMS | CfAA | WoT, SM, IoP | ISP, MBDP |
| SoA | BPS, DPP | IoS | KDoA |
| DIS | PHN | SoS | RtD |

The cascading hierarchy design in Table 2 represents the first step in the conceptual framework design in Fig. 2. The similarities between Table 2 and Fig. 2 are clear. The differences between the new understanding of artificial intelligence in CPS in Fig. 2 are also clear and very distinguishable from the existing understanding of artificial intelligence in CPS as seen in Fig. 1. Our approach for building the conceptual framework (Fig. 2) is based on an extensive review of literature that included multiple systems, models, and methodologies from over 90 leading articles on this topic. Concepts that reappeared in multiple articles were selected as the most prominent, and the relationships were recorded from each article. This enables a new approach to building the conceptual framework, based on complex socio-economic, organisation goals and policy issues that were identified in over 90 leading articles in this field, published in the past decade.

The taxonomy of abbreviations in Table 2 was derived from the taxonomy of literature in Table 1, which categorises the emerging concepts into a structure for artificial cognition in CPS. The structure relates the cognition in CPS with IIoT, bringing together the IoP and IoS, along with the process and transaction of IoT data. For example, the IoT data from DIS (see Tables 1 and 2 for definitions of abbreviations) connected to IoP and IoS, (representing systems of systems) enhance the cyber risk avoidance with real-time distribution and feedback directly from users and markets.

Thus, the evolution of cognition in CPS space adds a new perspective to the existing cyber risk avoidance mechanisms. The inter-relationships between these elements are crucial for defining dynamic cyber risk analytics with real-time probabilistic data. The current approaches taken for cyber risk analytics assume development of IoP and IoS and reliability of IoT. A deeper understanding of the relationship between IoT and I4.0, following the categories presented in Table 1, is required to develop a dynamic cyber risk analytics structure.

Furthermore, Table 2 shows that cognitive CPS capabilities are related to the integration of cyber physical capabilities into the industrial value chains. Hence, the proposed structure for cognitive CPS uses principles of IoT and integrates network intelligence, providing convergence, orchestration, and visibility across otherwise disparate systems. The proposed cognitive CPS also provides a structure for the operation and management of multiple CPS-related elements in the context of I4.0. Figure 2 shows the inter-relationship between different cognitive CPS communities, processes, societies, and platforms. The integration of cyber physical capabilities into the cognitive CPS involves the integration of IoT, WoT, SM, IoP, and IoS into SoS. With the use of Grounded Theory and Pugh-controlled convergence, the categories (from Table 1) are correlated in a hierarchical framework in Fig. 2, and correspond with the taxonomy (in Table 2). These are established models for decomposing and reverse engineering design processes. The hierarchical cascading in Fig. 2 explores the potential for automated and semi-automated methods that could be applied to ascertain and accelerate (and start to automate) the evolution of autonomous artificial intelligence in CPS. The concepts and the hierarchical structure in Fig. 2 originate from the summary map and the hierarchical cascading of actions in Table 1. The abbreviations are present in Table 1, and categorisation of taxonomic imperatives is first presented in Table 2. In Fig. 2, we apply the findings to build a conceptual diagram for visualising the updated CPS architecture.

The conceptual diagram in Fig. 2 originates from the re-evaluated five levels in the original 5C architecture, but without the 'cyber' level—described in Fig. 1. The remaining four levels are updated with concepts emerging from recent literature on this topic, with a time span between 2010 and 2020. We identified from literature (in the summary map Table 1) new and emerging concepts related to CPS architecture that are not included in the current 5C architecture. In Fig. 2, we present a hierarchical integration of these new and emerging concepts and present an updated 4C architecture—four levels of CPS architecture.

Since this review paper is built upon the notion of updating the existing 5C architecture in Fig. 1, we used the same conceptual order, but we integrated the improvements found in recent literature. We identified a lot of new terms, springing up between 2010 and 2020, and we wanted to put them in conceptual order. We anticipated this to be a required first step and a real service to future studies aiming to build a diagnostic architecture. The conceptual diagram in Fig. 2 derives new understanding on why cognitive evolution in cyber physical systems is inevitable and autonomous with the increased integration of connected devices (IoT). The hierarchical cascading in Fig. 2 is designed using the grounded theory approach for relating emerging concepts. The emerging concepts identified in the literature review are first presented in the summary maps, and then taxonomic approach is used to relate the categories and organise in a hierarchy of most closely to most distantly related. Conceptual design is then used to cascade the hierarchy in a framework. The framework in Fig. 2 explores how automated and semi-automated methods are accelerating (starting to automate) the evolution of autonomous artificial intelligence in CPSs. The framework in Fig. 2 represents a new mechanism and prototype of a hierarchical structure that facilitates deeper understanding of interconnected concepts—both being crucial given that there is no direct reference in literature to artificial cognition in CPS and cyber risk analytics.

**Fig. 3** Emerging CPS architecture—4 levels

| | | | | |
|---|---|---|---|---|
| Cognitive communities | 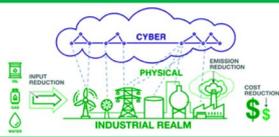 | | CPS, IoE, 5C, AoA, OoA, VOA, VEO, VEP, MDMS, SoA, DIS | Self-configure |
| Cognitive processes | 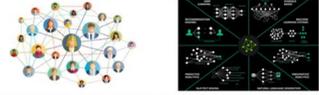 | | CDN, CfAA, BPS, DPP, PHN | Self-aware |
| Cognitive societies | 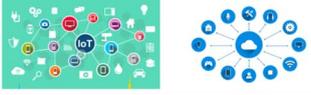 | | IoT, WoT, SM, IoP, IoS, SoS | Self-compare |
| Cognitive platforms | 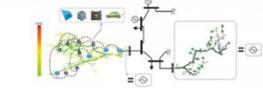 | | IPv6, ISP, MBDP, KDoA, RtD | Self-optimise |

For final comparison, we present a visualisation in Fig. 3 that compares the emerging four levels of CPS architecture. In Fig. 3, we can see that CPS as a concepts has evolved significantly since the 5 levels CPS architecture in Fig. 1.

In Fig. 3, we can see how the taxonomy from the summary map in Table 1 has been integrated in the four levels CPS architecture. We can also compare Figs. 1 and 3 to visualise the differences between the CPS architectures from 2015 and 2020. Although Fig. 2 presents the same information, in a conceptual diagram, we designed Fig. 3 for easier comparison.

# 7 Discussion

The updated four levels—CPS architecture in Fig. 2 offers a new and important step in updating our understanding of how CPS operate in 2020. Since the existing 5 levels CPS architecture (see Fig. 1) is few years old, and there has been many changes in connected systems since its creation, we considered this update timely and of relevance. We also argue that with the rise in new IoT and IIoT, complex, coupled and connected systems, such updates should occur at much faster intervals. This paper adopted the argument that AI should *'be programmed with a virtual consciousness and conscience'* (Meissner 2020), because we are in the middle of a new AI revolution that is changing our economy and society. There are studies investigating whether AI can create *'novel though'* (Fazi 2019). The mechanism in this paper describing how AI is evolving into CPS is based on grouping of future and present techniques and presenting the design process through a new hierarchical cascading design for a conceptual framework.

The conceptual framework in Fig. 2 details significant advancements over the past 5 years that can be seen in the most closely related framework on this topic in Fig. 1. For example, cognition in Fig. 1 is based solely on decision support system for prioritising workload, with a single focus on industrial processes. The new conceptual framework presented in Fig. 2 includes social machines, connected devices, and knowledge developments, among new concepts such as internet of services and internet of people.

The differences between the new framework in Fig. 2 and the earlier framework as seen in Fig. 1 mean that AI is evolving at a much faster rate than industrial understanding of this process. The new framework in Fig. 2 captures the changes in connected devices generating vast amounts of data, captured and stored in different heterogenous formats (e.g. high-dimensional data, real-time data, translytic data, spatiotemporal data). The new framework in Fig. 2 details the process of how the new data are captured, stored, processed, analysed, and used in near real-time, with low-latency. This is a very different process than our past understanding of CPS cognitive decision-making tasks, as seen in Fig. 1.

The main point of discussion from the new conceptual framework is that CPS are capable of much more than we describe in existing frameworks on CPS cognition in Fig. 1. With the availability of new types of data from IoT devices, CPS are becoming more automated. For example, with the new translytic data, CPS can transact and analyse data. With spatiotemporal data, CPS can map the demand in real-time. And with the complexities of high-dimensional data, CPS can understand the relationships between seemingly unrelated events and create new services and products. These new data streams are highly complex, and only AI can analyse such data and derive predictions with low-latency. Hence the evolution of AI in CPS is inevitable, autonomous, and it is already happening.

The arguments presented in this research are focused on understanding how the increased computational power of connected devices, has created intrusive self-building CPS, that represent human-like performance, triggering the creation of collective intelligence. The aggregated knowledge synthesised from recent literature, created a more comprehensive understanding of the current evolution of AI in CPS. We should not wait another 5 years before a new framework is designed to explain how AI is evolving with the emergence of new data formats, analysed with increased computational powers in connected devices.

# 8 Conclusion

In this paper, we have produced a hierarchical cascading framework for analysing the evolution of AI decision-making in cyber physical systems. The significance of the new framework is the findings that (1) such evolution is autonomous because of the increased integration of connected devices (IoT) in cyber physical systems; (2) such evolution is inevitable, because only AI can analyse the volume of data generated in low-latency, near real-time, hence, only AI can create value from new and emerging forms of big data. Nevertheless, we argue that the main value of the new 4 levels of CPS architecture, is the perception of cyber-physical systems as physical and human networks, where cognition emerges from the cyber-physical 'societies' and 'communities' (see Table 1). Our interpretation of CPS architecture perceives cyber-physical systems as social machines, and we place value in human interaction with such systems. In previous CPS architecture (Fig. 1), we can see that human intervention is predominated in the configuration level and the CPS depend on human cognition and there is a separate layer for 'cyber'. In our 4 levels of CPS, we integrated the 'cyber' in all levels, and we argue that there is a value for artificial intelligence to learn from human–computer interactions. Instead of relying only on feedback from connected devices, in some scenarios, human input is of much greater value. We have seen this in the current efforts to monitor a fast spreading pandemic—Covid-19. All contact tracing apps are based on human–computer input. Relying on computer data alone, was considered too slow and ineffective. We use this as a final example to rationalise our argument for perceiving future cyber-physical systems operating as social machines, obtaining input from humans and connected devices.

To present this rationale, in the taxonomic methodology, we adapted the summary map method, for transparency and justifications for concept selection and we used the literature review for decisions on the design of the hierarchical cascading. Out attempt in this paper was to contribute to the cultural practice of engineering discussion, on how artificial intelligence is evolving over time, by presenting a snapshot in time on this topic. It is evident that our analysis builds on considerable assumptions and guesses based on findings and taxonomic categorisation of existing literature. Our discussion in this paper points to the importance of understanding the impact of connecting complex and coupled systems.

Complex interconnected and coupled systems can evolve automatically with the continuous technological upgrades in existing CPS. The new hierarchical cascading framework in this paper identifies approaches to model imperative mechanisms within complex interconnected and coupled systems. In important environments for AI, such as IoT, we can model the connections and interdependencies between components to both external and internal IoT services and CPS in summary map. The summary map identifies the imperative categories for the evolution of artificial cognition in CPS. By applying established engineering design models, the summary map is advanced in a hierarchical structure for artificial intelligence in CPS. However, more empirical and philosophical research is needed on this topic before we can argue a comprehensive understanding on how artificial intelligence is evolving behind complex interconnected and coupled systems.